# Towards Empowering Diabetic Patients: A perspective on self-management in the context of a group-based education program

**Atae Rezaei Aghdam**
School of Information Systems
Queensland University of Technology (QUT)
Brisbane, Australia
Phone: +61 731384825
Email: atae.rezaeiaghdam@hdr.qut.edu.au

**Jason Watson**
School of Information Systems
Queensland University of Technology (QUT)
Brisbane, Australia
Phone: +61 731381656
Email: ja.watson@qut.edu.au

**Shah Jahan Miah**
Newcastle Business School
The University of Newcastle
Newcastle, NSW, Australia
Phone: +61 2 498 54043
Email: shah.miah@newcastle.edu.au

**Cynthia Cliff**
Faculty of Health
Queensland University of Technology (QUT)
Brisbane, Australia
Phone: +61 731383391
Email: c.cliff@qut.edu.au

## Abstract

With the exponential rise in chronic disease, Online Health Communities (OHCs) offer opportunities for people to share information and experiences. Type-2 diabetes is one of the most prevalent chronic diseases globally and one that can have a devastating effect on an individuals' life. This paper aims to investigate the potential for OHCs practices to empower diabetic patients who are participating in a face-to-face Diabetes Group Education Program (DGEP). Using a qualitative content analysis of the three most popular type-2 diabetes communities on Reddit, we identified three salient themes including (1) exchange lifestyle-related advice, (2) experience of commonality, and (3) brainstorm potential solutions for daily challenges. Our findings revealed OHCs could extend the value of the face-to-face DGEP by leveraging online value co-creation behaviour. This paper provides a novel framework for maximising the effectiveness of the DGEP and identifies avenues for future research.

## 1  Introduction

Diabetes is recognised as the world's fastest-growing chronic disease (Australia 2020; Lovic et al. 2020). According to the International Diabetes Federation (IDF), by 2040 one adult in ten will have diabetes (642 million) (Australia 2020). Diabetes is a chronic and progressive disease, which needs continuing self-management and self-awareness for a lifestyle change (Kjellsdotter et al. 2020). Self-management is one of the most key success factors impacting the progression of type-2 diabetes for patients, as the decisions that they make daily considerably impact their health outcomes (Funnell and Anderson 2004). Patients play a pivotal role in their self-care as they are doing more than 95% of their diabetes care outside of medical centres or at home (Su et al. 2019). OHCs as affordable and easily accessible 24/7 services, can facilitate self-management of diabetics by offering health-related advice and stories, social and emotional support (Aghdam et al. 2018). An OHC refers to a group of people who interact with each other in an online platform about similar health issues (Wang et al. 2017). Due to the fact that people tend to trust others who are in a similar situation rather than organisations, businesses, or government figures and media, it stands to reason that the content shared by peers in OHCs has potential to encourage community members to engage in health-related online activities (Irshad et al. 2013; Litchman and Edelman 2019). Participating in OHCs progressively transforms patients from passive recipients of healthcare services to active agents (Bragazzi 2013). As active agents, patients can access, share and integrate their resources, sharing their experiences and stories, and emotionally supporting peers to achieve their health-related goals (Forouzandeh and Aghdam 2019) (Aghdam et al. 2020). The empowerment of patients improves the patients' role in co-creation, co-designing, and co-delivering health services (Ciasullo et al. 2017). This is essentially a reality for people with chronic disease such as diabetes patients who need informational and emotional supports that allow them to be successful in their disease self-management (Litchman and Edelman 2019). In this regard, the diabetes education program has been a focus of prior research as a specific intervention that supports diabetes self-management (Findlay-White et al. 2020). OHCs provide opportunities for members to exchange new ideas, knowledge and information about diabetes self-management, functioning as a bridge among people with type-2 diabetes and healthcare professionals and providing online discussion platforms to brainstorm potential solutions (Sim et al. 2008). As such, this study aims to investigate the potential practices of online diabetes communities to address the following question; how can an online diabetes community empower patients in context of a Diabetes Group Education Program (DGEP)?

The remainder of this paper is organised as follows; the next section describes the background of the literature. The following section explains the research methodology followed by the trustworthiness process. The discussion section provides a comprehensive overview of the contributions of the study from both theoretical and practical perspectives and the final section synthesises the findings and provides avenues for future research.

## 2  Background

Chronic disease is generally of long duration, slow progression, and impacts the quality of life (Martz et al. 2007). The care for people with chronic diseases such as type-2 diabetes is often complex and requires self-management as an essential element of the chronic care model (Wagner et al. 2001). Self-management includes actions and behaviours to manage the psychical, emotional, and social effects of the chronic disease (Adams et al. 2004). One of the key methods for self-management of chronic disease and improved health outcomes is patient education (Ellis et al. 2004; Mensing and Norris 2003). Patient education is the keystone of chronic disease self-management and is significant in achieving positive health outcomes for chronic disease patients (Ellis et al. 2004; Mensing and Norris 2003). Patients need support, education, guidance and empowerment from their healthcare providers to tackle barriers to effective self-management (Diabetes 2009). Studies contended that participation in self-management courses also improves patient confidence, self-management skills and ability to self-manage their chronic disease, and improves the quality of life (Turner et al. 2015). Diabetes group education programs typically assist patients to achieve knowledge and skills and confidence to manage their diabetes as well as opportunities to interact with peers and healthcare providers (Jonkman et al. 2016). Group interactions facilitate further learning and raise motivation by interacting and learning from the experience of others (Odgers-Jewell et al. 2017). Research shows that diabetes group-based education programs benefit patients who derive social and emotional support from discussion with others (Steinsbekk et al. 2012). This type of active participation by patients in their health journey leads to value co-creation (Osei-Frimpong et al. 2015). In the healthcare context value co-creation refers to "activities centered around the individual patients or in collaboration with numbers of the service delivery network including the patient, family, friends, other patients, health professionals and the outside community" (McColl-Kennedy et al. 2012, p. 6). Value is co-created synergistically and digital health platforms such as OHCs act as a coordinating device between community members (Smedlund 2016). Thus, digital health platforms such as OHCs are ideal places for value co-creation (Aghdam et al. 2020; Kamalpour et al. 2020). Because of the nature of the DGEP, patients face three different stages during their health journey; (1) prior to joining, (2) during the program, and (3) after the program. We adapted Customer-Dominant Logic (CDL) to divide the customer journey into these three phases. In fact, CDL argues that customers control the service situation and control is a relevant issue in many domains, and due to progressively empowered customers (Seybold 2001), this direction will most likely continue in the future (Heinonen and Strandvik 2015). Accordingly, in the healthcare domain, empowering patients in OHCs can activate value co-creation behaviour among stakeholders (Litchman et al. 2018).

As diabetes self-management requires a patient-centred approach (Funnell et al. 2007), in particular for a demand-driven decision making (e.g. in clinical settings - Miah, 2013) and to date, the most successful diabetes self-management group activities and classes have been evaluated based on empowerment theory (Heisler 2010), our study applied empowerment theory in the context of OHCs and a diabetes group education program. Empowerment theory contends that actions, activities or structures might be empowering and the outcome of such process leads to being empowered (Zimmerman 2000). According to empowerment theory, people need opportunities to become active in community decision-making to improve their quality of life. As such, we leverage an affordable and easily accessible 24/7 digital health platform such as OHCs to facilitate the process of self-management through informational, social, and emotional support. The proposed framework goes beyond the existing system-centric approaches to a new mode of conceptualisation and practice, which focuses on interactions among all stakeholders in OHCs. The proposed framework addresses diabetes-related needs and challenges including; informational, social, and psychological needs.

## 3   Research Methodology

In this study, we selected online Reddit diabetes communities as our data source. The interactions between users are mostly focused on the posts themselves and members will give the post all their attentions. There are numerous diabetes-related topics in this forum, which make it a promising source of users' interactions for this research study.

### 3.1   Diabetes Online Communities on Reddit

Reddit is a social aggregation and public discussion website. In Reddit, three popular diabetes communities comprise of more than 60,000 members. In this study, we selected *r/diabetes, r/type2diabetes,* and *r/diabetes_t2* communities. Within each community, there are a variety of threads and topics discussed by users. The total number of users in all of these communities was 59,400 in April 2020. From each topic, the tile and the content (e.g., textual information) were extracted without the additional information of the authors. A total of 189 topics were collected from Reddit from October 2019 to April 2020. In total 1989 threads were collected for analysis.

### 3.2   10-week Diabetes Group Education Program (DGEP)

Research argued that participation in diabetes group education program has multiple benefits for patients such as social and emotional support, and sharing experiences (Odgers-Jewell et al. 2017). The Queensland University of Technology (QUT) offers a partnership face-to-face DGEP to patients who are newly diagnosed or living with the type-2 diabetes long term. Over the course of program, the QUT DGEP aims to provide a quality lifestyle intervention empowering type-2 diabetic patients to better manage their symptoms. The DGEP runs for 10 weeks and includes various types of activities such as; initial assessment, weekly one-hour personalized exercise session, one-hour interactive group education regarding diabetes-related topics (e.g., diet, mindfulness, foot care, living with a chronic condition, etc.), and a final assessment at the end of the program. The outcomes of this award-winning program are promising and all patients involved no longer needing to stay on the long waiting list of the hospitals. One of the most important objectives of the DGEP is to keep patients connected whilst outside the program. We, therefore, aim to extend the value of the face-to-face DGEP by identifying the potential practices on the online diabetes communities, proposing a diabetes OHC framework for keeping patients connected to the program after discharge from the program.

### 3.2.1 Data collection

We collected data from the three popular Reddit diabetes online communities (r/diabetes, r/type2diabetes, and r/diabetes_t2). Reddit is a popular forum for diabetes (Duggan and Smith 2013). There are numerous health-related topics on this website, which make it a promising source of users' interactions for this research study. In addition, Reddit allows researchers to mine its data. Hence, we used the Python Reddit Application Programming Interface (API) Wrapper (PRAW) to collect the data. PRAW is a Python package that allows researchers to access, parse topics and subreddit, and extract the associated reply threads. As inclusion criteria for selecting posts and threads, we selected type-2 diabetes-related topics with more than 10 replies on each topic to obtain enough information. The interactions between users are mostly focused on the posts themselves and members will give the post all their attention. Demographic information about the participants was anonymized to guarantee the confidentiality and privacy of participants' data. In every stage of this research study, we followed the code of ethics for researchers of the Queensland University of Technology (QUT). The approval number is 1900001024.

### 3.2.2 Data analysis approach

In this study, we conducted an inductive thematic analysis to identify emergent themes from the data. The six steps of thematic analysis provided by (Clarke et al. 2015), guided us to identify the salient themes. Following the six steps of the thematic analysis and with the assistance of the NVivo 12 qualitative analysis software, we manually generated an initial list of codes. During the first step, we performed an initial analysis of the relevant topics and threads and recorded our notes via *memo* and *annotation* features of NVivo 12. In the second step, we inductively generated 106 nodes. In the third step, we combined codes revealing three overarching themes and nine subthemes. In the fourth step, which was reviewing and refining the themes, we reviewed all themes and subthemes to make sure that they followed a coherence pattern. During this phase, two themes were integrated because of their common content. In the fifth step, we concisely named the identified themes to reflect the story behind each theme and reflect what the themes are about. Hence, we named themes that address the research questions. Finally, in the sixth step, findings were synthesised to provide a concise and coherent report. In terms of testing the trustworthiness of the findings, we employed percent agreement as our method of inter-coder reliability checking. Two scholars, experienced in qualitative research and thematic analysis, checked different parts from creating initial codes to naming the themes. Each of them independently analysed the entire data and during the first meeting, the per cent agreement was 75% and after the second meeting, discussing the essence of the themes, a consensus was achieved and the overall results were 100%, making us confident about the reliability of our findings.

## 4 Findings

After performing the thematic analysis, our analysis resulted in three emergent themes from the data. Themes include (1) exchange lifestyle-related advice, (2) experience of commonality, and (3) brainstorm potential solutions for daily challenges. Table 1, summarises the thematic analysis outcomes. As evidenced in Table 1, patient participation in diabetes online communities leads to the co-creation of value. For instance, in theme 1, patients shared resources such as articles, and videos with peers. Another key finding is to improve patients' psychological wellbeing by participating in online activities such as story sharing and encourage other members of the community in self-monitoring.

| No | Theme | subtheme | example quotations for themes |
|----|-------|----------|-------------------------------|
| 1 | Exchange life-related advice | - Experience sharing<br>- Resource exchange<br>- Sharing new technologies benefits | ***Reddit User 1***. *"Is it safe to have sugar free foods/drinks?"*<br>***Reddit User 2***. *"I don't use artificial sweeteners, so I can't speak from experience, but I found this video interesting (Video URL)"*<br>***Reddit User 1***. *"Oh that was interesting thank you very much"* |
| 2 | Experience of commonality | - Share same situation<br>- Feeling less isolated<br>- Encourage peers in self-monitoring and self-care | ***Reddit User 3:*** *"I feel super motivated to bring down my diabetes"*<br>***Reddit User 4:*** *"Find you passions in life and live them, whatever they maybe.*<br>***Reddit User 3:*** *"Do it despite diabetes. Do it to kick diabetes' ass!"* |
| 3 | Brainstorm potential solutions for daily challenges | - Sleep problems of diabetics<br>- Challenges of diabetics at workplace<br>- Carrying medical equipment for diabetics | ***Reddit User 1***. *"How can I handle carrying diabetes bag for some places such as theatre or public and private businesses when the security ask me to not bring my bag into these places?"*<br>***Reddit User 2***. *"Perhaps a clear bag works so they could see what is inside"* |

*Table 1. Summary of the thematic analysis outcomes*

## 4.1 Exchange lifestyle-related advice

OHCs provide an opportunity for users to enhance their knowledge about symptoms, share their experience and advice. Information sharing by peers, experience and advice sharing, life-style related advice sharing, and sharing daily-basis activities are the most common activities identified by researchers in this study through thematically analysing the content of threads in the Reddit diabetes online communities. These types of giving and seeking (exchanging) advice are illustrative examples of value co-creation behaviours. In the value co-creation process, stakeholders such as organizations, patients or caregivers share, integrate and renew each other's resources (Palumbo et al. 2017). Our analysis shows that diabetic users mostly shared their experience, stories, and online resources (e.g., research articles, YouTube videos, and websites' URLs). Resource exchange is a mutual action taken by stakeholders in OHCs to access, share, and integrate resources (Beirão et al. 2017). In this regard, OHCs can facilitate resource exchange among stakeholder. For instance, members of the diabetes communities shared their up-to-date information and experience about using wearable devices (e.g., Continues Glucose Monitors (CGM), Dexcom and sport watches) for self-monitoring their health condition and reduce the burden of living with diabetes and improve quality of life.

## 4.2 Experience of commonality

The Experience of commonality in OHCs provides opportunities for members to feel that they are not alone. Hence, the experience of commonality is associated with positive mental health, improving emotional wellbeing Members of the diabetes communities perceived these online platforms as great places to tackle the feeling of loneliness and isolation. Especially in the current situation of the global outbreak of COVID-19, these online communities are ideal places to tackle psychological distress and depression. At this particular point in time, diabetic patients need to strengthen their sense of community by connecting and supporting each other in the OHCs. Because of the nature of OHCs, which provides access to information and coordinated social interactions, the members of these communities benefit an alternative solution and needs such as improving their wellbeing (Zhao et al. 2015). Emotional support directly impacts on the ability to self-manage diabetes and equally self-management of diabetes influences emotional wellbeing (Schiøtz et al. 2012). Sharing the same situation and stories with other members is another aspect of emotional support. Sharing the same stories creates a shared sense of meaning and community for users. In the Reddit diabetes communities, a large number of users encourage peers in their self-management of diabetes. Patients also reported that sharing monitoring data such as blood glucose and weight makes them feel empowered and motivated. Members of these communities' support each other in coping with social and emotional barriers, staying motivated to reach their goals, and encourage better self-care habits without fear of judgement or stigma. Improving the emotional wellbeing of diabetes leads to better self-care, overcoming psychological barriers, and ultimately, a better quality of life.

### 4.3 Brainstorm potential solutions for daily challenges

OHCs are ideal places for brainstorming solutions by members. We identified brainstorming of potential solutions to address daily challenges as another co-creation behaviour occurred in diabetes online communities. Virtual brainstorming is one of the most significant benefits of OHCs for diabetic patients. It provides an opportunity for community members to contribute new ideas to address diabetic daily challenges such as carrying medical equipment, diabetic's workplace problems, injection, and sleep problems. This was identified in many threads posted by the members of the communities. Hence, OHCs are ideal places to brainstorm potential solutions to address these issues. As it can be viewed in Table 1, Reddit patient suggests a solution to another patient, who is struggling with carrying diabetes bag in public and private business areas. These types of solutions are another example of value co-creation behaviour within diabetes online communities. Participating of community members in brainstorming activities, make them feel that their contributions are valuable and their ideas will help peers to tackle some daily challenges.

### 4.4 Proposed framework

Because of the high number of demands for DGEP, patients need to be in a waiting list before joining the program. While they are in a waiting list, they can communicate with discharged patients and use their experience. The resource exchange help patients reduce their stress and better prepare for the program. During the program patients, share their experiences and health-related stories with peers, encouraging each other to reach their health-related goals. During this phase, OHC can play an important role as an online interactive platform to facilitate patient-to-patient and patient-to-HCP interaction. After discharge from the program, patients still need to stick to their plans and self-manage their diabetes. OHCs provide opportunities for them to keep connected to the program, interact with HCPs and share their experience of the program with patients, who are in the "prior-to-joining" phase. Figure 1, demonstrates the proposed framework for diabetes online communities.

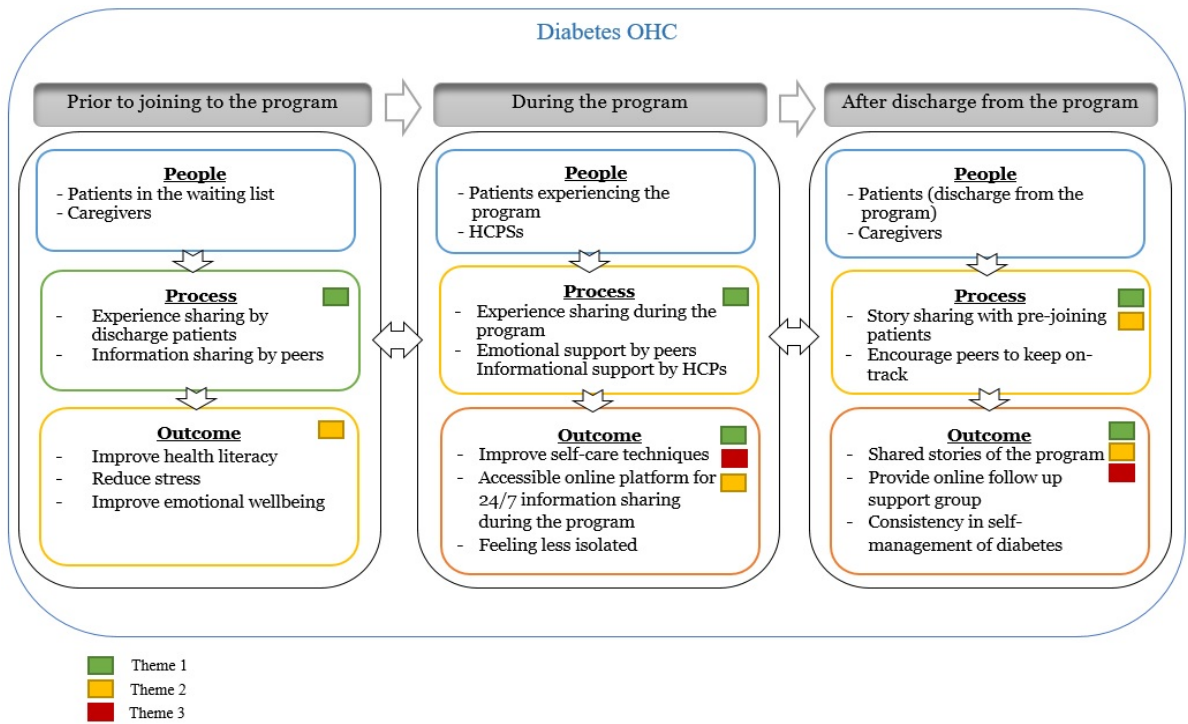

*Figure 1. The proposed framework for diabetes online communities*

In terms of the validity and utility of the proposed framework, we conducted expert interviews. A qualitative assessment of the framework flexibility was carried out through interviews with healthcare professionals and the diabetes program coordinator at the QUT clinic. Two health experts, who are directly involved and organised the DGEP, and have more than ten years of experience in the healthcare domain initially evaluated the framework and provided their feedback and suggestions.

## 5 Discussion

OHCs are proper educational platforms that lead to better health outcomes and members can learn more from others on how to better manage their health conditions (Chen et al. 2019). Information that shared by patients in OHCs benefits other patients by learning from peers, improving their self-management of disease, and ultimately, improving their health (Yan and Tan 2014). The proposed framework focused on patients as active agents in the process of online value co-creation. Patients are pivotal stakeholders in our framework that can co-create value by resource exchange and social support. In our framework, diabetic patients participate in different forms of value co-creation through informational, social, and emotional support. OHCs empower patients to actively engage in co-creation activities 24/7 especially in times of fear, isolation, and uncertainty. This research study has been conducted during the global pandemic of the COVID-19. During this pandemic and isolation time, patients increasingly participate in diabetes online communities to gain and offer emotional support. These easy-access and 24/7 online platforms help patients to tackle psychological issues such as depression, anxiety, and loneliness because diabetic patients have a twofold greater risk of depression (Schram et al. 2009). Shared stories and experiences in OHCs make patients feel that they are not alone, strengthening their sense of community by connecting and supporting each other.

### 5.1 The connection between themes and the face-to-face DGEP

Digital health platforms such as diabetes online communities have the potential to increase easy access to diabetes self-management interventions and techniques in the lower cost (Rosal et al. 2014). Furthermore, participating in diabetes group-based education program provides opportunities for patients to meet and discuss with other members of the communities, obtaining social and emotional support (Steinsbekk et al. 2012). In doing so, in recent years, the Queensland University of Technology (QUT) offer a partnership program to patients who are newly diagnosed or living with the condition long term. This program aims to provide a quality lifestyle intervention and empowering Type-2 diabetic patients to better manage their symptoms. As type-2 diabetes is a self-managed disease, one of the main aims of the program is to keep patients connected to the program after discharge. We adopted Customer-Dominant logic (CDL) as a way to explore overlaps between our findings and the DGEP. CDL is focused on activities and experiences of the customer at three different stages: pre-service, service, and post-service (Heinonen and Strandvik 2015). It is used as a way to extend the customers' perceptions of the offering and to extend market interactions (Heinonen and Strandvik 2015). Following CDL, we divided the diabetic patient journey into three phases including; (1) prior to joining the program, (2) during the program, and (3) after discharge from the program. In each phase, we identified value co-creation behaviours such as; resource sharing, story and advice sharing, and social and emotional support. Theoretically, we extended the target body of the knowledge in the healthcare service delivery through enhancing the empowerment theory in which, patients are the central facet and healthcare professionals and healthcare organisations are facilitators of the value co-creation process (Funnell and Anderson 2004). Research studies have overlooked the nuances relationship between empowerment theory, value co-creation, and the role of OHCs as facilitators for this process. This study provides an opportunity for leveraging peer-to-peer support within digital health platforms such as OHCs to empower patients in their self-management of diabetes. Practically, our findings further provide recommendations to the healthcare industry on how to effectively contribute to the online intervention by shifting from traditional dyadic interaction between healthcare professionals and the patient to online co-creation among all stakeholders. We believe that healthcare providers can potentially use our theoretical and empirical findings to extend the value of the face-to-face diabetes group-based education programs by keep patients connected to the program 24/7 regardless of their geographical distance with lower cost.

## 5.2  Limitations and directions for future works

Our study is not without limitations, yet these limitations provide interesting avenues for future research. Our data were gathered from Reddit diabetes online communities. We selected three popular diabetes communities on the Reddit to analyse the contents and interactions among members. We might overlook some small communities related to type-2 diabetes. Furthermore, we only used Reddit as our data collection source. Future studies can focus on more diabetes online communities, aiming that how can a fully functional assistive artefact be designed for diabetic patients, using the design science research guideline (Miah 2008; Miah et al. 2019). Although the face-to-face diabetes education program held in Queensland, Australia, it can be generalised to any other organisational or country context (For example, in decision support implementation (Ali, Miah and Khan, 2018) ensuring empowering end users). Our future study will extend the current framework by conducting interviews with the members of the communities to identifying their current level of engagement with OHC, identify benefits and challenges of using these platforms, and investigate their online value co-creation behaviour. Therefore, there are some areas required for further research. Another future avenue is to investigate the perspective in which healthcare organisations indirectly participate in online value co-creation. Experimental design studies of OHCs to explore the behavioural and psychological aspects of social support could also be useful.

# 6  Conclusion

In this research study, we sought to extend the current understanding of the potential of diabetes online communities in empowering self-management for diabetic patients. As such, the main aim of this study was to investigate the potential practices of online diabetes communities to empower self-management of diabetic patients in their health journey. Findings show that patients in diabetes online communities share information, experiences, stories, and potential solutions. They actively participate in online activities regarding offering and receiving support from peers. The vast majority of the shared contents on diabetes online communities include lifestyle-related advice such as diet, exercise and using wearable technologies to better monitor and care of diabetes. Members of diabetes online communities contend that these online forums are ideal platforms to obtain social and emotional support from peers. Our findings, which investigated the connection between Diabetes online communities' practices and outcomes and the real-world DGEP case can further assist healthcare organisations to effectively contribute to the online intervention and extend their communication channel from a traditional power balance between HCO and patients to interactive platform that enables all stakeholders to actively engage in value co-creation activities. As discussed, type-2 diabetes is a chronic disease that needs ongoing self-care and self-manage. OHCs provide opportunities for them to encourage each other in regards to sticking to their self-monitoring and self-management. This is especially true when they discharge from the DGEP and have no access to face-to-face interactions.

Zimmerman, M. A. 2000. "Empowerment Theory," in *Handbook of Community Psychology*. Springer, pp. 43-63.